# Emerging Optical Microscopy Techniques for Electrochemistry

Jean-François Lemineur,[a] Hui Wang,[b] Wei Wang,[b,*] Frédéric Kanoufi[a,*]

[a] Université de Paris, ITODYS, CNRS, 75006 Paris, France.

[b] State Key Laboratory of Analytical Chemistry for Life Science, School of Chemistry and Chemical Engineering, Nanjing University, Nanjing 210023, China.

[*] corresponding authors, e-mails: frederic.kanoufi@u-paris.fr, wei.wang@nju.edu.cn

**ABSTRACT**

An optical microscope is probably the most intuitive, simple and commonly used instrument to observe objects and discuss behaviors through images. Although the idea of imaging electrochemical processes *operando* by optical microscopy was initiated 40 years ago, it was not until significant progress made in the last two decades in advanced optical microscopy or plasmonics that it could become a mainstream electroanalytical strategy. This review illustrates the potential of different optical microscopies to visualize and quantify local electrochemical processes with unprecedented temporal and spatial resolution (below the diffraction limit), up to the single object level with subnanoparticle or single molecule sensitivity. Developed through optically and electrochemically active model systems, optical microscopy is now shifting to materials and configurations focused on real-world electrochemical applications.

**KEYWORDS**

Optical microscopy, electroanalysis, electrochemical conversion, single entity electrochemistry, nanoparticles, *operando* imaging, spatiotemporal resolution

# 1. INTRODUCTION

Electrochemistry is a vivid branch of science, particularly within the search for renewable energy sources and systems enabling the conversion and storage of energy. While great efforts have been made toward the synthesis and processing of electroactive and electrocatalytic materials, often emphasizing the importance of their structuring at the nanoscale, the improvement of the performance of most electrochemical devices is hampered by the kinetic limitations of electrochemical reactions. The understanding of their mechanisms and fundamentals relies on the establishment of structure-function relationships, particularly at the nanoscale. This has then driven the shift of traditional electroanalytical strategies and techniques based on ensemble-averaged methods, *e.g.*, current-potential, response toward the imaging of electrochemical processes with higher sensitivity, spatial and temporal resolution and manyfold complementary information.

Despite considerable progress in advanced *in situ*/*operando* characterization techniques, optical microscopy remains the only technique that requires simple operating procedures while being noninvasive and enabling multiple instrumental couplings. Optical imaging of electrochemical processes was introduced in the mid-1980s (1) along with scanning electrochemical microscopy. It was not until the improvement of optical detectors (and components) and the development of plasmonics that electroanalytical strategies employing optical microscopy were brought back to the fore. The recent concepts of superlocalization, allowing to track phenomena with a resolution of a few nanometers which is lower than the smallest picture element, *i.e.*, the pixel, open many perspectives for imaging electrochemical processes beyond the diffraction limit. Several reviews have detailed the general operating principles and applications of such advanced optical microscopies in (nano)chemistry, sometimes in electrochemistry (2–5). Herein, we summarize their recent achievements in the imaging of multifarious electrochemical systems. After a short description of some of the microscopes used, we show how they are currently employed to resolve and quantify the heterogeneity of electrochemical interfaces, from the macroscopic scale to the single nanoobject or even to the subentity or single molecule level.

## 2. OPTICAL MICROSCOPES

A detailed description of the operating principle and configurations of the various optical microscopes employed in electrochemistry can be found in (2–5). This review mostly focuses on the use of wide-field microscopes, in which the light emanating from a whole substrate is collected by a microscope objective and captured in a single snapshot by a camera or a spectrograph for spectroscopic imaging. They offer higher spatiotemporal resolution imaging than point scanning, *e.g.*, confocal or tip-enhanced Raman scattering, microscopes: within a single >50x50µm$^2$ image snapshot, within millisecond timescale, thousands of localized electrochemical behaviors can be simultaneously obtained.

These microscopes are sensitive to the optical properties of the sample of interest,

mostly absorbance, refractive index, scattering or luminescence. The growing popularity of optical microscopy approaches in electrochemistry is not only related to the ability to image but also the collection of quantitative information from the mathematical treatment of the optical signal; see, *e.g.*, (3). Plasmonic metals, *e.g.*, Au or Ag, constitute a highly sensitive detection tool in optics, as their interaction with light induces the surface-confined oscillation of their free electrons known as surface plasmon resonance (SPR). The SPR is strongly sensitive to the metal local charge density or the refractive index of its environment, enabling different plasmonic-based imaging of electrochemical processes. In SPR microscopy, the light locally reflected by the interface between a thin layer of Au (used as an electrode) and an electrolyte produces an SPR image sensitive to a wide variety of (electro)chemical reactions (2, 6). Localized SPRs, or LSPRs, are supported by plasmonic nanoparticles, NPs, or nanostructures. Tracking the scattering (or LSPR) spectra of single plasmonic NPs allows sensing single NP electrochemistry (2, 4, 7). The illumination of plasmonic NPs or nanostructures (roughened electrodes) also produces a strong electromagnetic near field able to enhance the Raman scattering generated by (individual) molecules by several orders of magnitude (8, 9). The local increase in Raman intensity is used to provide molecular vibrational images in surface-enhanced Raman scattering and SERS microscopy with single molecule sensitivity.

Bright field and reflectivity microscopes mostly use (axial) illumination along the objective axis and collect light transmitted or reflected by the sample of interest. They can probe absorbance, refractive index or light emission, such as fluorescence or Raman scattering, after appropriate filtering of the excitation light beam. Oblique incidence illuminations are used mostly for (i) dark-field configurations and (ii) extreme total internal reflection (TIR) conditions, as they offer a lower optical background level, enabling single nanoparticle or single molecule imaging sensitivity.

Dark-field illumination avoids blurring the detector, which only collects the light scattered by the sample of interest. It has mostly been used to image the scattering of plasmonic NPs with the eventual spectroscopic capture of their LSPR spectrum. However, a broader class of scattering NPs has more recently been imaged at higher sensitivity by interferometric scattering microscopes (10).

The TIR condition allows confined illumination (by evanescent waves) to light up only objects located within a few hundred nm above the illuminated interface, which is particularly useful for single-molecule fluorescence detection. Similar TIR illumination conditions are used in plasmonic-based (SPR) microscopy.

Imaging without optical illumination, and therefore at the lowest optical background, is possible using the electrochemical triggering of a luminescent reaction, named electrochemiluminescence microscopy (5, 11, 12). As it involves chemically unstable precursors of the luminescence reaction, electrochemiluminescence offers chemically confined illumination of objects near the electrode and single photon sensitivity (13).

Finally, a microscope is characterized by two crucial notions: its sensitivity, *i.e.*, its ability to detect an object, and its resolution, *i.e.*, its ability to distinguish two objects close together. Microscopes are diffraction limited, meaning that objects should be separated by a distance greater than λ/2NA, with λ being the illumination wavelength and NA being the objective numerical aperture. Furthermore, single objects smaller than this limit, *e.g.*, single NPs or single molecules, appear in an optical image as an identical optical pattern, regardless of their structure or composition, named the point spread function (PSF) or Airy disk. Note that the resolution of localization of optical microscopes can be greatly improved by image posttreatment consisting of approximating the PSF by a two-dimensional Gaussian distribution and algorithmically extracting the spatial origin of a single emitter, also named its optical center of mass or centroid. By superlocalization approaches, the location of various electrochemical reactions is visualized *operando* at single nanoentities with a resolution <5 nm.

**3. OBJECTS OF STUDY**

**3.1 IMAGING OF HETEROGENEOUS INTERFACES**

Probing nanoscale electrochemical events at heterogeneous interfaces discloses the internal mechanism and detailed dynamics of electron transfer processes in analytical chemistry and biosensing. Different imaging strategies have been developed to visualize local electrochemical information at electrode interfaces, such as plasmonic, electrochemiluminescence, and fluorescence microscopy (1–8). These revolutionary studies reveal the intrinsic characteristics and mechanisms of nonfaradaic and redox processes with ultrasensitive temporal and spatial resolution. This section details the optical electrochemical imaging of various heterogeneous interfaces.

**3.1.1 Individual cells**

Monitoring single-cell responses to substrates or small molecules and cellular processes at the microscopic level deepens the understanding of the mechanisms of physiological and biochemical dynamics. Optical techniques have been introduced to study multifarious single-cell electron transfer events with high spatial resolution, providing detailed information on their transient activities and local distributions (19, 20). Tao et al. first developed plasmonic-based electrochemical impedance microscopy to uncover heterogeneous processes such as single-cell apoptosis and electroporation with millisecond time resolution (21). A local impedance measurement ($Z$) is derived from the local change in plasmonic intensity ($\Delta\theta$) of a thin Au SPR surface according to $Z^{-1}(x, y, \omega) = j\omega\alpha\Delta\theta(x, y, \omega)/\Delta V$, where $\omega$ is the angular frequency of the AC modulation, $x$ and $y$ are the locations on the electrode, and $\alpha$ is a constant determined by theoretical calculation or experimental calibration. They optically resolved the local impedance of subcellular structures and ion distributions in mammalian cells with submicrometer spatial resolution and excellent detection sensitivity (~2 pS). By combining the electrochemical plasmonic impedance imaging method with the traditional patch clamp technique (Figure 1a), the fast propagation of the action potential in individual mammalian neurons was mapped without any labeling (22). They further investigated the heterogeneous distribution of ion channels at the subcellular level and proposed studying various cellular electrochemical activities and understanding the related biological functions and mechanisms.

*INSERT FIGURE 1*

Recently, the groups of Sojic and Paolucci developed a surface-confined microscope based on electrochemiluminescence illumination of objects and illustrated it to map membrane adhesion sites of single cells on an electrode (20, 23). Their groups further demonstrated the influence of photobleaching on electrochemiluminescence emission. As both photo- and electrochemical activation involve the same excited state, the more photoactivated the fluorophore is, the less active it is in the electrochemiluminescence. Despite this issue, new imaging strategies combining fluorescence recovery and electrochemiluminescence were envisioned (24).

**3.1.2 Fingerprints**

Visualizing latent fingerprints (LFPs) is an essential method for biometric identity authentication. Various chemical and physical strategies have been explored to reveal LFPs, including multimetal immunodeposition, fluorescence, and ink staining (25–30). The ability of electrochemical techniques to identify explosive residues and other chemicals secreted by LFPs has gradually gained attention. Tao et al. demonstrated a plasmonic imaging technique combined with electrochemical measurement to map human LFPs on an electrode surface (28). Finger secretions block the electron transfer process on the electrode, and the plasmonic contrast of the local fingerprint region is transposed into a local electrochemical current of redox-active molecules in solution. Later, Su and coauthors (Figure 1b) developed an electrochemiluminescence-based imaging technique to enhance and visualize local LFPs using different dyes: Ru(bpy)$_3^{2+}$, rubrene, and electropolymerized luminol (25, 27, 29). The electrochemiluminescence signal was generated only between the ridges of the LFPs, and different details of the LFPs were resolved: the bifurcation, core, island, pore, lake, peak, and termini.

Recently, Hu et al. reported a new strategy for transferring and imaging LFPs onto nonporous substrates using simultaneous interfacial separation of a polydopamine film and electroless silver deposition (27). As sweat components and underlying substrates were well preserved, they generalized the approach to different substrates, regardless of surface hydrophobicity or micromorphology.

### 3.1.3 Bipolar electrochemistry

A bipolar electrode (BPE) is a conductive material exposed to an external electric field from the application of a potential difference between two electrodes in an electrolyte. The potential difference induces electrical polarization at opposite poles of the bipolar electrode, which manifests as a gradient in the distribution of free electron density (31). When the potential difference is large enough, opposite electrochemical reactions occur simultaneously at both ends of the BPE. As these reactions occur without external current flow, their demonstration requires complementary visualizations, such as probe-labeled imaging with electrochemiluminescence reagents, pH chromogens, or fluorescent dyes, and label-free plasmonic imaging techniques (32–34). Crooks and coauthors developed electrochemiluminescence-based imaging of BPE reactions. They provide a means to locally quantify the thermodynamics and kinetics of the reactions to spatially reconstruct the voltammogram of these reactions from an electrochemiluminescence image. A triangular BPE is used, which allows, for electroanalysis, the quantification of the reaction of interest on the smallest part of the BPE (toward the point of the triangle) compared to the larger counterelectrode reaction part (35).

Xu and Chen reported an ultrasensitive wireless electrochemiluminescence biosensor for quantitative monitoring of c-Myc target mRNA in tumor cells on a BPE substrate (36). In this system-on-chip, they integrated RuSi@Ru(bpy)$_3^{2+}$ for optical signal amplification with a 24-fold improvement over Ru(bpy)$_3^{2+}$-NHS labels. Beyond electrochemiluminescence, Kuhn et al. successively presented other indirect imageries of BPEs based on pH-triggered local precipitation (37) or fluorescence modulation (32). The products of these reactions are monitored in the vicinity of the BPE.

Except for the above techniques using optical probes, plasmonic-based microscopy provides label-free visualization and thus there is no need to engage a faradaic reaction at the BPE a priori. Wang and coauthors first demonstrated the capability of plasmonic imaging to directly visualize the interfacial potential distribution on a bipolar microelectrode array with a sensitivity of 10 mV (33). The external electric field controls the redistribution of the free electron density on the BPE and thus modifies its local dielectric (optical) properties. The local plasmonic response is thus predicted using the Drude model. Furthermore, it is possible to locate the zero-potential line on BPEs, where no reaction occurs, regardless of their geometry (*e.g.*, round, triangular, hexagonal, star, and diamond shapes) during nonfaradaic charging and faradaic processes (38). The results revealed that the geometry of the electrode and the nature and redox potential of the faradaic reactions affect the position of the zero-potential line on the BPE.

**3.1.4 Two-dimensional nanomaterials**

Two-dimensional (2D) nanomaterials are emerging as novel platforms for optoelectronics and biosensing due to their unique physical, chemical, and electronic characteristics (39–41). The spatial charge distribution of these thin layers has been studied by optical techniques, such as plasmonic, bright field, or interference scattering microscopy, coupled with electrochemical measurements (6, 41–43). The optical readout reveals fundamental electrochemical parameters of 2D electrodes and their heterogeneity, such as quantum capacitance and local charge density, with high spatial and temporal resolution. Graphene is the most studied 2D material without a bandgap. Tao and coauthors mapped local electron and hole puddles with charged impurities in a graphene monolayer by plasmon-based impedance microscopy (42). The surface charge density, $\Delta\sigma$, is related to the local interfacial capacity per unit area ($c$) and the applied potential ($\Delta V$), according to $\Delta\sigma = c \cdot \Delta V$. Periodic modulations of the potential control the surface charge, resulting in a modulation of the plasmonic intensity ($\Delta\theta$). From the latter $\Delta\theta$, extracted from the Fourier transform of the graphene region images, a local capacitance distribution is obtained according to $c \sim \Delta\theta/\Delta V$. Further charging induces an expansion of the graphene according to Pauli repulsion. This expansion is imaged using a nm-sensitive optical edge-tracking method (44). The technique allows determining the electromechanical stress that increases quadratically with the modulation of the applied potential and extracting the Young's modulus of different regions. Further oxidation of graphene at potentials > 1.4 V results in its conversion to graphene oxide. This process was imaged *in situ* by a label-free refractive index-sensitive optical technique such as interference reflection microscopy (IRM). This reveals the formation of flower-like patterns from which the local degree of graphene oxidation can be quantified and its chemical vs. electrochemical oxidation compared (43).

Apart from graphene, molybdenum disulfide ($MoS_2$) monolayers are another attractive 2D material for next-generation nanoelectronic devices, with a direct bandgap of 1.9 eV. Tao et al. imaged the local charge distribution of atomically thin $MoS_2$ upon electrochemical charging. The change in charge induces a local change in the absorption of $MoS_2$, which is imaged by bright-field transmission microscopy with higher sensitivity (45).

### 3.1.5 Electrocorrosion and electrodeposition at a large interface

Electrocorrosion and electrodeposition are classical strategies for fabricating functional interfaces and improving the surface characteristics of metallic, ceramic, or polymeric materials (46, 47). The basic principle of electrocorrosion and electrodeposition is the destruction and formation of materials on a working electrode immersed in an electrolyte solution and subjected to an external potential. The morphological evolution of the electrode surface during these interfacial engineering processes is essential to uncover the detailed dynamics and accurately determine structure-function relationships. Different optical imaging techniques have been used to probe corrosion processes, *e.g.*, fluorescence, reflectivity or confocal microscopy. They allow the identification of locally different reaction products or solution pH or the identification and measurement of the size of crevices. V. Pérez-Herranz developed a wide field reflectivity microscope allowing real-time observation at the scale of several cm$^2$ on copper and stainless steel electrode surfaces (48). They also mapped the different corrosion behaviors of crevices and grain boundaries and identified the generation of gas bubbles. Smyrl et al. used fluorescence microscopy to image the regions where oxides, which trap fluorophores, preferentially form at higher resolution. The measurement is complemented by confocal measurements allowing a topographic (3D) image of crevices (49, 50). Vivier and coauthors proposed a quantitative assessment by reflectivity imaging of the thickness of passive layers during corrosion of carbon steel under polarization (51). They performed local reflectivity measurements of the steel surface during cathodic and anodic polarizations to study the formation of $Fe_2O_3$ and $Fe_3O_4$ oxides. It allows the identification of the most active regions of the surface and the establishment, simultaneously over each µm² regions of the mm² imaged surface, of local voltammograms of their activity or transformation.

These same imaging techniques are also used to follow *operando* the electrodeposition processes at micrometric scales, in particular in energy storage or conversion systems (52), to identify the formation of dendrites (53, 54) localized *operando* at nanometer resolution (Figure 1c) or electrode passivation (55) and to remedy them.

## 3.2 SINGLE NANO-ENTITY STUDIES

The growing use of nanoscale objects is bound to the identification of their intrinsic properties, for which quantitative nanostructure-activity relationships are urgently needed. In electrochemistry, different cross-reading approaches have been proposed at the single NP level (56–59), mainly based on their isolation in time, one electrochemical event at a time, or in space, by local electrochemical probing. In addition to probing local electrochemistry with nanoelectrodes or nanopipettes, in the so-called scanning electrochemical (SECM), electrochemical cell (SECCM) or ion conductance (SICM) microscopies configurations, one can use optical microscopies that allow high-sensitivity imaging and detection at the resolution of a single NP. Recent developments proposed to integrate optical microscopy readout with such scanning electrochemical microscopes (60, 61). The coupling of optical microscopy with SECCM is of particular interest to image the subtle electrochemical processes occurring inside the nano- or microelectrochemical cell, obtained by the confinement of a droplet of electrolyte by a nano- or micropipette. Indeed, in addition to high resolution electrochemical imaging, SECCM allows a high throughput exploration of local electrochemical processes by a versatile modifications or benchmarking in each droplet of experimental parameters, *e.g.* surface or solution composition or electrochemical interrogation. While SECCM enables nanoscale electrochemical exploration with single droplet resolution, optical microscopy affords a complementary subdroplet imaging resolution.

Coupling electrochemistry to optical microscopies appears relevant to probe *operando* nanoscale electroactivities. Optical movies allow high-throughput readout of individual NPs within large ensembles, altogether submitted to the same experimental condition, allowing identification of subpopulation behaviors, for example by drawing and comparing their individual electrochemical activity (*e.g.* current-potential, charge-time, etc. curves). Moreover, NPs can be differentiated by their size, structure or composition from their different optical properties, typically their optical cross section (related to their refractive index, absorption, luminescence, etc.).

In the field of NP electrochemistry, optical microscopy has been applied to reveal and study a wide variety of chemical or physical processes illustrated in Figure 2a for the particular case of Ag-based NPs, *i.e.*, the most studied system due to their plasmonic activity and easily tunable (electro)chemical activity.

*INSERT FIGURE 2*

### 3.2.1 Electrodeposition and electrodissolution

Optical monitoring of an electrode surface during electrodeposition reactions allows simultaneous capture of the moment and location of formation of individual NPs with a very large field of view (up to millions of NPs simultaneously (62)) and camera temporal resolution (up to >1 kHz). Optical imaging then provides statistically significant data to test and enrich NP nucleation/growth mechanisms and models.

For noble metals, *e.g.*, Ag, the differences in the onset of NP appearance on the electrode reflect the variability of their nucleation barrier (63, 64). Iron group metals also reveal competition with other electrode reactions, such as water reduction in the case of Ni or Co (65, 66). Beyond the local chemical information, localization of the nucleation sites (67, 68) allowed reconstruction of each diffusion zone around the NPs and probing diffusion cross-talk between neighboring sites.

Within a region of interest (ROI) centered on each NP, the transient evolution of the local optical intensity is extracted from optical movies. Such transient gives insights into the modes and kinetics of single-NP growth (69). This can be converted into the amount of locally electrodeposited material (63, 64, 70) due to a calibration between NP size and optical intensity obtained from the optical images of gauge NPs, *ex situ* correlative SEM analysis or optical modeling. Combined with Faraday's law, local currents, in the form of optovoltammograms (Figure 2b), associated with the growth/dissolution of each NP are obtained by this quantitative analysis, again evaluated for hundreds of NPs simultaneously.

The reverse electrodissolution reaction was also studied for electrodeposited NPs (63, 64, 71) or nanocolloids immobilized on an electrode (69, 72, 73). The disappearance of the optical feature associated with the NP in the images accounts for its dissolution dynamics, investigated for metallic Ag NPs in different electrolytes (63, 64, 71).

Electrodeposition/stripping is an attractive strategy to decorate electrodes with a high density of NPs of controlled size distribution, allowing the examination of structure-activity relationships. The effect of NP size on their oxidation potentials, observed for Ag NPs, validates Plieth's theory that links the electrochemical stability of NPs (below ~50 nm) to their surface tension (63, 64).

The electrodissolution of Brownian nanocolloids was also probed by optical microscopy. It allows tracking the motion of Brownian NPs in solution during their collision (reactive or not) with a polarized electrode and complements electrochemical nanoimpact experiments (56), in which the current spikes associated with a reactive collision of NPs provide information on their size, dispersion, stability, concentration, etc. The correlated optical and electrochemical detections revealed a more complex picture. 3D optical tracking of Ag NPs near a polarized interface revealed intermittent NP-electrode interactions associated with partial oxidation events (72, 74). This supports the hypothesis of multistep $Ag \rightarrow Ag^+$ electrodissolution, first established from high-frequency current traces and attributed to stochastic electrical disconnection (75). Under conditions favoring $Ag^+$ precipitation, a time lag is observed between the injection of electrochemical charges and the dissolution of optically probed NPs, highlighting solid phase conversion (*e.g.*, to oxide, halide, or thiocyanate crystals (72, 73, 76–78)).

This type of optical imaging of the appearance or disappearance of optical features, primarily used with model metal NPs, was extended to visualize and quantify *in situ* the formation or dissolution of a variety of other materials, such as gas nanobubbles (79–83) or ionic crystals (84), and holds promise for high-throughput monitoring of structural deformations of nanoelectrocatalysts under operating conditions (85). The technique can be easily extended to the study of various phase formation processes as long as they can be triggered electrochemically, either by direct electrodeposition or indirectly by local electrolysis.

### 3.2.2 Electrochemical conversion

During the electrochemical or redox conversion of an NP, the change in the redox state is often associated with a change in its optical properties, such as fluorescence, absorption, or scattering cross-section. Different optical microscopies can distinguish the initial and final states of redox conversion of individual NPs. Gradual changes in composition can even be monitored *in situ* and in real time, revealing mechanistic pathways at the single NP level (Ag examples in Figure 2a).

The conversion of Ag-based metallic NPs into $Ag^+$ salt nanocrystals is evidenced under dark-field microscopy by a decrease in the intensity of the light they scatter without, however, reaching total signal extinction. A spectrometer inserted at the end of the optical path, or a hyperspectral camera, completed the pure optical imaging with their UV-vis spectrum, providing information on the composition and conversion mechanisms in solutions of $Ag^+$ precipitating agents (77, 78, 86–88).

The transformation of metallic NPs, *e.g.*, Ag, into more noble metal NPs, *e.g.*, Au, by galvanic replacement, a popular redox reaction in colloidal synthesis, was followed *in situ* under $Au^{3+}$ solution flow by dark-field microscopy. The optical transients are also characterized by a sudden drop in the scattering signal but observed after a waiting time of variable duration. The broad distribution of waiting times confirms the gradual transformation of solutions. The difference between single vs. ensemble NP behaviors suggests that the transformation is kinetically limited by the stochastic formation of a void in the NP lattice (broad distribution). Once the void is formed, the NP transformation is rapid (sudden drop) and diffusion-limited (89). Other mechanistic indications were identified, such as the role of precipitating $Cl^-$ or the NP ligand shell (90, 91).

The methodology is applicable to nanomaterials used for energy storage or conversion. The refractive index of $LiCoO_2$ NPs decreases linearly with the amount of Li-expelled ions, allowing imaging of their electrochemical (de)lithiation by refractive index-sensitive microscopies (92). From the optical intensity fluctuations of individual NPs recorded during cyclic lithiation/delithiation voltammetry or nanoimpact experiments (93), Wang and coauthors quantified the dynamics of Li-ion diffusion with an optically inferred current sensitivity of 50 fA.

Similarly, for supercapacitor applications, the insertion/deinsertion of alkali ions into electrochromic NPs, such as $WO_3$ or Prussian blue (PB), was imaged under bright-field transmission (94–98). The evolution of the light transmitted by individual NPs (Figure 2c) analyzed according to the Beer-Lambert law allows quantification of their conversion rate. For some $WO_3$ NPs, slower and less complete insertion dynamics, even more pronounced for NP aggregates, were observed, which suggests irreversible trapping of $Li^+$ at the NP-NP or NP-electrode interfaces. The heterogeneity of ionic nanocrystal-electrode contacts has also been highlighted when potassium ions are inserted into electrochromic PB nanocubes (99). Sputtering an ultrathin layer of Pt onto electrode materials, as is often done in SEM, reconnects and renders all nanocubes electroactive and avoids erroneous conclusions in establishing structure-activity relationships.

**3.2.3 Electrocatalytic systems**

*3.2.3.1. Probing molecular intermediates*

Quantification of the electrocatalytic activity of single NPs is achieved by probing the molecular products or intermediates of the reaction by fluorescence microscopy or surface-excited Raman spectroscopy (SERS), sometimes with single-molecule sensitivity (see Section 2.4). These microscopies mainly use a redox molecular probe (commonly phenoxazine dyes such as resorufin), in which one of the redox forms is luminescent or Raman active, or use pH-sensitive probes.

By adapting the strategy developed for photocatalysis (100), fluorescence microscopy allowed imaging and evaluating (i) the deactivation of Pt/C electrocatalysts during the hydrogen oxidation reaction (101) or (ii) the 2- vs. 4-electron reduction pathways of $O_2$ by magnetite NPs (102).

Willets et al. imaged by SERS the local activity and distribution of reaction potentials on Ag NP aggregates for the 2-electron conversion of Nile blue (103, 104). A recent work suggests possible extension to the direct detection of valuable reaction products such as $CO_2$ and its reduction products (105).

Electrochemiluminescence involves redox and electrochemical reactions that can be activated by Au NPs (106), resulting in NP visualization. Under the oxidizing conditions of electrochemiluminescence, the reaction is quickly deactivated (fading image) owing to Au oxide formation, which was prevented using Janus Au-Pt NPs (107).

*3.2.3.2. Probing entities transformation*

Owing to the sensitivity of the LSPR of a metallic NP to its free-electron density, Mulvaney et al. (108) proposed monitoring the shift in LSPR wavelength by spectroscopic scattering microscopies to image and quantify the flow of (few) electrons during (dis)charging of Au NPs. The method has since been used to probe any (electro)chemical reaction that would perturb the electron density of NPs (109, 110) to evaluate the rate of oxidation of ascorbic acid by $O_2$ (111).

The LSPR is also influenced by the refractive index of the chemical environment of the NPs, which allows imaging the electrocatalytic reactions that modify it, as illustrated by Long et al. (112) during the oxidation of $H_2O_2$ on Au nanorods. The nanorod surface, first oxidized to Au hydroxide/oxide, is then reduced back to Au while oxidizing $H_2O_2$, with the two activities represented by different LSPR shifts.

Local refractive index changes following electrocatalytic reactions at nonplasmonic NPs can also be detected, but they must be deposited on an optically active electrode that allows such sensitivity, *e.g.*, in plasmonic-based or interference scattering microscopy. The concept, developed by Tao et al. (28), to obtain local electrochemical activities of heterogeneous electrodes allowed the establishment of hydrogen evolution reaction (HER) CV at single Pt NPs (113).

Electrocatalysis of the HER or oxygen evolution reaction (OER), which is critically important in energy applications, often leads to the formation of gas bubbles. Bubbles are thought to nucleate and grow in regions supersaturated by gas molecules. Optical microscopies that can probe bubble production at the micro (114) and nanoscale (115) enable the identification of the most active catalysts.

Nanobubbles (NBs) were revealed during HER on Au nanoplates by TIR fluorescence microscopy by tracking the adsorption of a single rhodamine molecule at the electrolyte-gas interface (Figure 3a). The collected fluorescence intensity additionally allows NB size estimation (79, 80). Due to their low refractive index, NBs are also directly detected by label-free microscopy (Figure 3b). Optically undetectable Au-Pt NPs were revealed from their electrogenerated NBs (116). Superresolution plasmonic-based (117) or interference-reflection (82) microscopy allows rapid dynamic mapping of NB nucleation sites on Au or ITO electrodes. The optical response further provides a dynamic estimation of the size and shape of NBs (or contact angle), suggesting that Pt nanocatalysts were rapidly electrically isolated by NBs (83), and their continuous growth proceeded through spill-over (79).

*INSERT FIGURE 3*

**3.2.4. Superlocalizing physical changes**

Controlling *operando* the deformation or structural alteration of NPs is crucial for the longevity of electrocatalytic energy conversion devices (85). Such information, often hidden in electrochemical analysis, except for some stochastic collision experiments (56), is within the reach of optical imaging via spatial superlocalization in 2D or 3D of the centroid of the optical pattern of NPs (118) during electrochemical solicitation.

The edge-tracking procedure suggested by Tao and coauthors allows localization of the contours and thus evaluation of an apparent size of objects with a size comparable or higher than the diffraction limit (119). The strategy highlighted the reversible breathing of single $Co(OH)_2$ particles while being electrochemically probed in the OER region (120).

The motion and orientation of individual electroactive pseudo-2D graphene microplatelets were optically tracked as they approached and collided with a microelectrode (Figure 3c). The latter yields variations in the overall electroactive surface area that correlate with transient variations in the electrochemical current (121) until they rearrange themselves flat on the electrode (122). The dynamics of the process are obtained from the rate of angular motion of the platelet.

## 3.3 PROBING SUBNANOENTITIES

Optical imaging can image beyond the single NP resolution and probe transformations at the subunit level. First, an NP is separated into two classical subunits: the shell, which is in contact with the outer environment, can exhibit a different reactivity than the NP core. With the help of dedicated optical models, the contribution of these subunits can be revealed optically. In a second approach, anisotropic or localized electrochemical processes inside an NP are revealed by a superlocalization approach (2.2.4).

### 3.3.1 Surface alteration

Imaging the shift in LSPR wavelength of plasmonic NPs by spectroscopic DFM allows probing (electro)chemical conversion of the NP shell. The strategy developed to monitor the deposition of Ag on strongly scattering Au nanostars (123) allows the detection of the underpotential deposition of Ag on various shaped Au nanocrystals. An LSPR shift of a few nm corresponds to submonolayer deposition. An optical voltammogram (Figure 4a) is obtained from the LSPR frequency variations, revealing the influence of the crystallographic facet orientation on the Ag electrodeposition potential (124). The electrochemical conversion of the Ag shell to AgCl was monitored in a similar manner (Figure 2a). The reaction intermediates distinguished optically from their different plasmonic coupling modes suggest propagation of the Ag/AgCl interface between the Au core and the chloride electrolyte interface (125).

The reversible electrochemical de/amalgamation of Au NPs by Hg (126, 127) was similarly imaged. It first involves saturation of the NP surface by Hg atoms before their slow diffusion into the core. The reverse process operates in the same way but with a slower solid diffusion rate for the expelled Hg atoms.

Similarly, Link, Landes et al. (110) distinguished the reversible physical adsorption of chloride ions on the surface of an Au NP from the irreversible formation of an Au chloride shell. Then, they generalized the method to probe the electrochemical adsorption dynamics of various molecules or anions (128).

As discussed in 2.2, the chemical reactivity of the NP surface, not restricted to plasmonic NPs, was probed using refractive index-based techniques. Plasmonic-based imaging has been used to differentiate between surface and bulk oxidation (or reduction) for Au NPs or electrodes (129, 130). Interferometric scattering microscopy has been more recently introduced to electrochemical studies, although it shows high imaging sensitivity of various charge transfer processes. Although at the $LiCoO_2$ microparticle but in a real Li-ion battery configuration, it allowed *operando* dynamic imaging of local Li ion flow during (des)insertion (Figure 4b), revealing how its heterogeneity can alter battery operation (131). It could also identify the restructuring of electrochemical double layers at ITO or Cr nanostructures in iodide electrolyte (132).

*INSERT FIGURE 4*

### 3.3.2 Centroid superlocalization

Edge-tracking procedures (119) were used to evaluate the electrochemical deformation of gold nanowires (133) and graphene sheets (44) due to surface stress and Pauli repulsion, respectively.

The superlocalization of AgCl NPs colliding with a cathodically biased electrode revealed conversion in multiple motion-reaction steps attributed to loose electrical connections (Figure 2a). Chloride ions are released locally in multiple steps, each creating a limited silver metal inclusion within the NP and propelling the NP to a nearby reactive site (86).

If the displacement of the NP PSF over distances greater than the NP dimension reveals their physical motion, a slight spatial fluctuation can be attributed to an asymmetric transformation of the NPs, highlighting the presence of inactive zones within the NP.

A shift in the centroid of Ag NPs was observed by Willets et al. during their oxidation (134), suggesting asymmetric dissolution limited by the asymmetric formation of a nonconductive surface oxide (Figure 2a).

Similarly, the reduction of single PB NPs (98) to Prussian white is not always complete. The position of the optical centroid depends on the intermediates formed locally and therefore fluctuates during the conversion (Figure 4c). A microscopy approach was then proposed to evaluate the propagation of the reaction along the vertical direction. Since no vertical propagation is detected, conversion is thought to occur via a shell-to-core model.

Ultimately, Wang and coauthors showed that Fourier transform analysis of the optical images enables pushing the superlocalization procedure down to subnanometer accuracy. By optically imaging the charge separation in Au nanorods subjected to periodic capacitive charge-discharge cycles, they detected a periodic subnanometer centroid shift (Figure 4d), suggesting heterogeneous charge accumulation on the Au surface (96). This unprecedented resolution should unlock the label-free observation of nanoscale local surface chemistry or the manipulation of single NP local reactive sites (135).

**3.4 SINGLE MOLECULE ELECTROCHEMISTRY**

Molecular electrochemistry is a very broad subject involving many new concepts and techniques, some with an unprecedented level of spatial resolution giving it a new impetus, such as for the establishment of structure-function relationships at the single-molecule scale (14, 15, 59, 136–138). In 1995, Fan and Bard first demonstrated a single-molecule electrochemical measurement. This uses the concept of current amplification by catalytic redox cycles, which involves repeating the oxidation and reduction events of a molecule placed between two electrodes (139). To date, high spatial resolution optical approaches have been developed to capture the intermediate states of the electrochemical reaction of a single electroactive molecule, such as surface-enhanced Raman spectroscopy and single-molecule fluorescence spectroscopy (14, 16, 140, 141). Very recently, these methods have been transposed to electrochemiluminescence imaging by Feng and coauthors (Figure 5a), showing how, in complete darkness but by precise control of the chemical reaction between electrogenerated reactants, electrodes can turn-on single-photon chemiluminescent reactions (13). This opens a fascinating area of molecular electrochemistry and electroanalytical research.

*INSERT FIGURE 5*

More abundant literature uses the former two optical approaches detailed here. The tiny optical response during the transition of different redox states of a single molecule can be followed to probe the electrochemical dynamics. This reveals the intrinsic mechanism of electron transfer reactions in homogeneous solutions, enabling the fundamental understanding of the electrochemistry of single molecules.

**3.4.1 Surface-enhanced Raman spectroscopy**

Single molecule surface-enhanced Raman spectroscopy (SM-SERS) can be used to directly probe individual heterogeneous electrochemical events in a single molecule (8, 14, 16, 138, 142). It provides fundamental information about structural changes and specific behavior of a surface reactive site with respect to a redox couple or to understand molecular electron transfer mechanisms and intracellular dynamics in analytical chemistry. Plasmonic nanostructures locally enhancing the Raman intensity are defined as "hot spots" in SERS in which vibrational information is captured to determine the redox transient states of target electroactive molecules.

In 2010, SM-SERS was applied for the first time to discover electrochemical events in a bianalyte system combining two dye molecules, rhodamine-6G (R6G) and Nile blue (143). In an open-frame electrochemical cell, the presence of distinct imprinting modes of a molecule at a hot spot was captured to address redox (on-off) events. In addition, a thought-provoking question was answered as to whether the "average" behavior of a bulk system can be recovered from the events of a single molecule. Identical local conditions can only be extracted from measurements of a single molecule without averaging electrical or optical properties. In parallel, Van Duyne implemented SM-SERS to study single electron transfer events ($O + 1e^- \rightleftarrows R$) of the dye molecule R6G adsorbed on a silver NP under nonaqueous conditions (144). The broad local distribution of reduction potentials can be attributed to variations in molecular orientations and variations in the local surface site or chemical potential of the R6G-Ag bonding units.

Recently, Wilson and Willets demonstrated the superresolution imaging strategy of SM-SERS with sub-10 nm accuracy by establishing the spatial relationship between the centroid of the SERS emission and the corresponding maximum intensity (104, 145, 146). Using this approach, they visualized the specific redox potentials at different adsorption sites of individual Nile blue molecules on colloidal Ag NPs. The reversible trajectories of the centroid of the molecules on the surface of the NPs during a redox cycle were attributed to the location-dependent potentials of the single electroactive molecule, where the SERS intensity modulates the activation and deactivation states with oxidation and reduction processes.

### 3.4.2 Single molecule fluorescence spectroscopy

The synergistic coupling of electrochemistry with single molecule fluorescence spectroscopy (SMFS), via confocal laser scanning, TIR, or superresolution microscopes (15, 59, 137, 147, 148), allows the study of heterogeneous electron transfer events by simultaneously capturing a quantitative electrochemical signal and *in situ* fluorescence images. The key feature of this coupling is to obtain both temporally and spatially resolved information by following the electron transfer process. The redox states of the electrofluorochromic compounds at the single-molecule limit can be determined from the blinking (on/off states) of the fluorescence response. As the electrode potential varies, the residence time constant in each of the states (on/off, then ox/red) reflects the rate of the redox transformation and thus the dynamics of the electrochemical reaction.

In 2006, Bard and Barbara demonstrated for the first time the possibility of studying single-molecule electron transfer processes by spectroelectrochemistry (149). They studied hole-injection oxidation events of single molecules of poly-9,9-dioctylfluorene-cobenzothiadiazole (F8BT), a redox conjugated organic polymer used in solar cells and flat panel displays, immobilized on an ITO electrode. As oxidation quenches fluorescence, the electron transfer dynamics are studied as a function of potential and illumination. If both the excited and ground states of F8BT can be oxidized, only the ground state oxidation shows a narrow distribution of fluorescence turn-off potential, revealing its half-wave potential.

The technique has been extended to the study of more conventional fluorescent probes in bioimaging. Gooding et al. were the first to report reversible fluorescence switching of bovine serum albumin (BSA)-conjugated Alexa Fluor 647 redox probes by TIRF (150). The potential-modulated fluorescence of BSA-Alexa Fluor 647 immobilized on ITO at the single protein level was studied by measuring the variation in the number and intensity of fluorescent spots. The observed pH dependence indicates the involvement of two-electron one-proton transfer in the fluorescence switching mechanism.

Orrit and coauthors (151) studied the fluorescent readout of redox-sensitive methylene blue probes at the single-molecule level enhanced by individual gold nanorods (Figure 5b). MB, a common redox indicator for tissue staining and biosensing, undergoes a reversible fluorescence change to form colorless methylene blue by two-electron one-proton transfer. Time traces of the fluorescence flashing of a single electrogenerated MB molecule are recorded at different potentials. The residence times in the on/off states are evaluated by a step detection algorithm. The distribution of these residence times at each potential is used to evaluate the half-wave potential of single molecule electrochemical switching from the Nernst equation.

## 4. PERSPECTIVES AND CONCLUSION

This review has shown how advanced optical microscopies are now able to image a wide range of electrochemical phenomena with unprecedented temporal and spatial resolutions (below the diffraction limit), up to the single object level with subentity, subnanoparticle or single molecule sensitivity. By providing quantitative descriptors complementary to electrochemical signals, they have unraveled old problems while revealing new ones in the different fields explored by electrochemistry (sensors, electroanalysis, corrosion, electrocrystalization, energy conversion and storage, electrocatalysis, etc.).

First developed through model systems, they are now shifting to materials and configurations focused on real-world applications, where they can be exploited to precisely locate heterogeneous electrochemical processes, distinguish domains (electrode regions, nanoobjects, etc.) of different structure/composition and therefore distinguish competing chemical routes, or identify the origin of problems to fix. As definitely the most intuitive platform to see *operando*, optical microscopies should become a routine electroanalytical tool to evaluate the performance of electroactive materials and rationalize their design or degradation. An even deeper degree of understanding can be reached from their simple implementation with complementary structural and chemical analyses such as spectroscopy (UV-vis or Raman, as well as the promising surface-enhanced IR) or within multicorrelative microscopies combining, *e.g.*, local electrochemical probes and *in situ* TEM. Particularly, approaches combining optical visualization within complementary electrochemical local probing, *e.g.* by SECCM (57, 60), will enable the generation of large sets of correlated optical and electrochemical data. It should become a powerful approach for benchmarking wide range of electrochemical situations.

The generalization of these explorations to broader electrochemical situations also implies seeing with greater sensitivity (*e.g.*, iSCAT (10, 132) or photothermal microscopes) more rapidly in more complex media (seeing through fog is within reach) or in real-world systems (optical fiber explorations). It is also necessary to generalize the nature of current collectors (optoelectrodes) providing sensitive optical detection ensuring homogeneous (electro)chemical contact with the objective of studying minimal electrocatalytic activity, for which transparent carbon- or graphene-based electrodes are promising. Finally, the thousands of data per image, even tenfold with complementary spectroscopic data, promise to unlock many structure-function understandings. The use of artificial intelligence will be crucial to achieve faster automated postprocessing, *e.g.*, object identification by deep learning (152), or recognition of electrochemical behavior and for the removal of unnecessary information to optimize data storage and processing in real time.

**ACKNOWLEDGMENTS**

F.K. acknowledges support from the European Union's Horizon 2020 Research and Innovation Programme under Marie Skłodowska-Curie MSCA-ITN Single-Entity Nanoelectrochemistry, SENTINEL [812398]. J.-F.L. and F.K. acknowledge the Université de Paris and CNRS for financial support. W. Wang and H. Wang acknowledge the National Natural Science Foundation of China (Grants 21925403, 21904062 and 21874070) and the Excellent Research Program of Nanjing University (Grant ZYJH004) for financial support.

**FIGURE CAPTIONS**

**Figure 1.** (a) Schematic illustration of the plasmonic-based electrochemical impedance imaging technique of action potentials in single neurons. A micropipette is patched on single neurons cultured on the surface to trigger action potentials, which are recorded by patch clamp electronics and plasmonic imaging. Adapted with permission from Reference (22). Copyright 2017, John Wiley & Sons. (b) Schematic illustration of a typical electrochemiluminescence imaging technique for visualizing the latent fingerprints on electrode surfaces with negative and positive modes. Adapted with permission from Reference (29). Copyright 2012, John Wiley & Sons. (c) Superlocalization of Zn dendrite nucleation and growth monitored by dark-field microscopy in a Zn aqueous battery configuration. Adapted with permission from Reference (54). Copyright 2021, Elsevier.

**Figure 2.** (a) Summary of optical microscopy studies reporting single silver-based NP electrochemistry grouped into three main categories: growth, dissolution and conversion, with corresponding references. Adapted with permission from Reference (3). Copyright 2021, John Wiley & Sons. (b) Quantitative light scattering monitoring of silver NP deposition and stripping voltammetry. The optical intensity transients (extracted in ROIs) are quantitatively converted into single NP currents and optovoltammograms. Adapted with permission from Reference (64). Copyright 2018, John Wiley & Sons. (c) Optical transmittance monitoring of $WO_3$ NP electrochemical conversion. The different optical transients reveal heterogeneous Li-ion insertion in single NPs and aggregates. Adapted with permission from Reference (94). Copyright 2019, American Chemical Society.

**Figure 3.** (a) Imaging of gas nanobubbles nucleating and growing upon electrocatalytic water splitting in the vicinity of four nanocatalysts by TIR fluorescence microscopy. Adapted with permission from Reference (79). Copyright 2018, National Academy of Sciences. (b) $H_2$ nanobubbles equivalently detected at single Pt nanocatalysts by interference reflection microscopy. The optical data are further exploited to estimate the evolution of the contact angle of single NBs during the growth process. Adapted with permission from Reference (83). Copyright 2021, American Chemical Society. (c) Optical tracking of the graphene platelet impact event and further dynamic rotation at a polarized microinterface. Adapted with permission from Reference (122). Copyright 2021, American Chemical Society.

**Figure 4.** *Operando* optical screening enables subentity studies. (a) Optically inferred voltammetry tracking facet-dependent underpotential deposition of silver atoms on single gold truncated octahedral nanocrystals. Adapted with permission from Reference (124). Copyright 2020, CC BY 4.0. (b) Optical image showing the front of the phase transition in the $Li_xCoO2$ cathode particle, from which the charge-discharge dynamics are revealed *operando* in a real Li-ion battery. Adapted with permission from Reference (131). Copyright 2021, Springer Nature. (c) The optical centroid

motion of a single Prussian blue NP during oxidation/reduction cycles reveals local transformations or inactive sites. Because the centroid is moving back to its initial position, the conversion is reversible. Adapted with permission from Reference (98). Copyright 2021, CC BY-NC 3.0. (d) Ultimate tracking resolution: the electrochemical charging of single Au nanorods results in subnanometer optical centroid motion owing to local electron accumulation. Reproduced with permission from Reference (153). Copyright 2019, American Chemical Society.

**Figure 5.** Electrochemistry with a single-molecule fluorescence readout. (a) Principle of single electrochemiluminescent event observation enabling single molecule luminescence imaging, without illumination, of arrays of nanoband electrodes. The dilution of both the dye and coreactant electrogenerated intermediate imposes a single reaction event, i.e. the formation of the excited dye molecule further emitting a single photon, located where the reaction was initiated during the image snapshot. Adapted with permission from Reference (13). Copyright 2021, Springer Nature. (b) Schematic showing the local fluorescence emission under plasmon-enhanced photoactivation of the electrogenerated fluoroactive form of a single immobilized electrofluorophore. The red/ox proportion of the single molecule is obtained from blinking (right part) of the fluorescence signal at a fixed electrode potential. Adapted with permission from Reference (151). Copyright 2021, John Wiley & Sons.

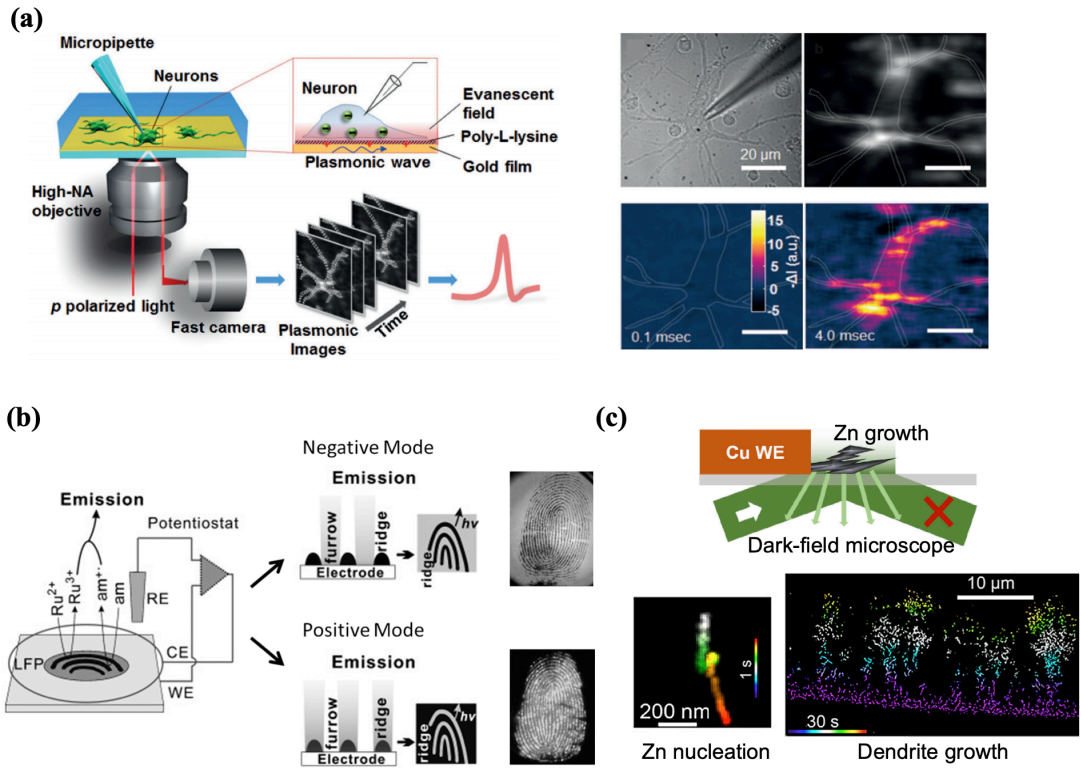

**FIGURE 1**

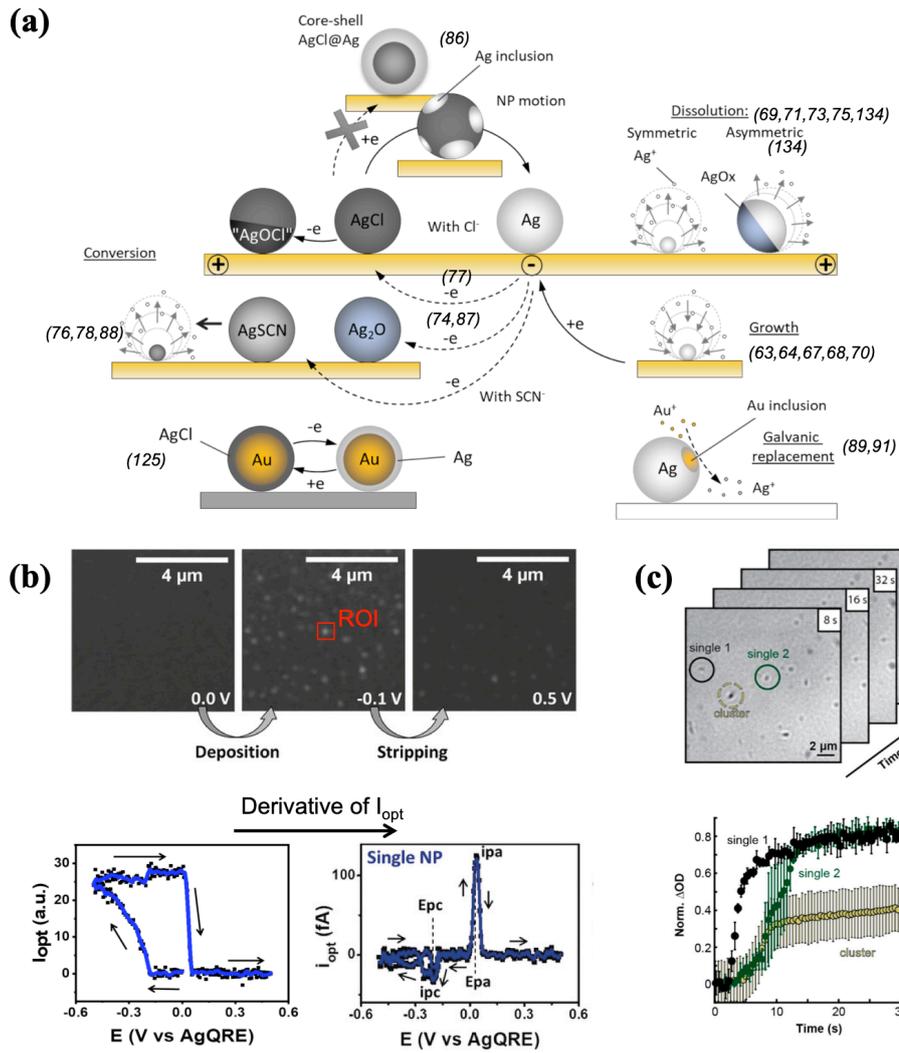

**FIGURE 2**
49

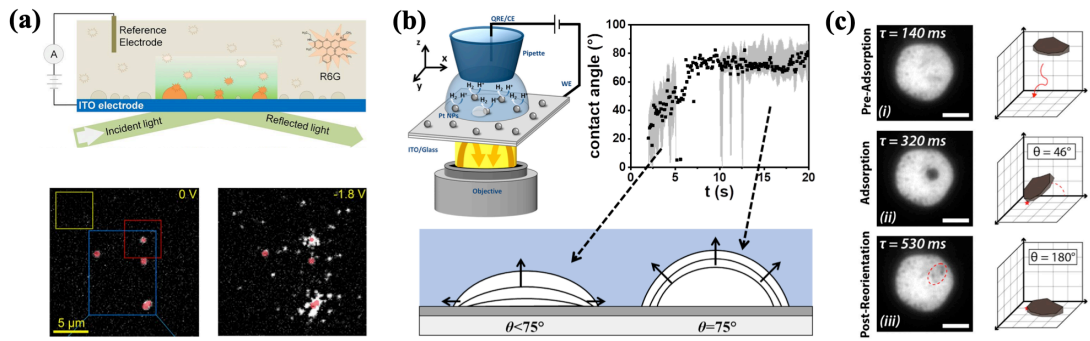

**FIGURE 3**

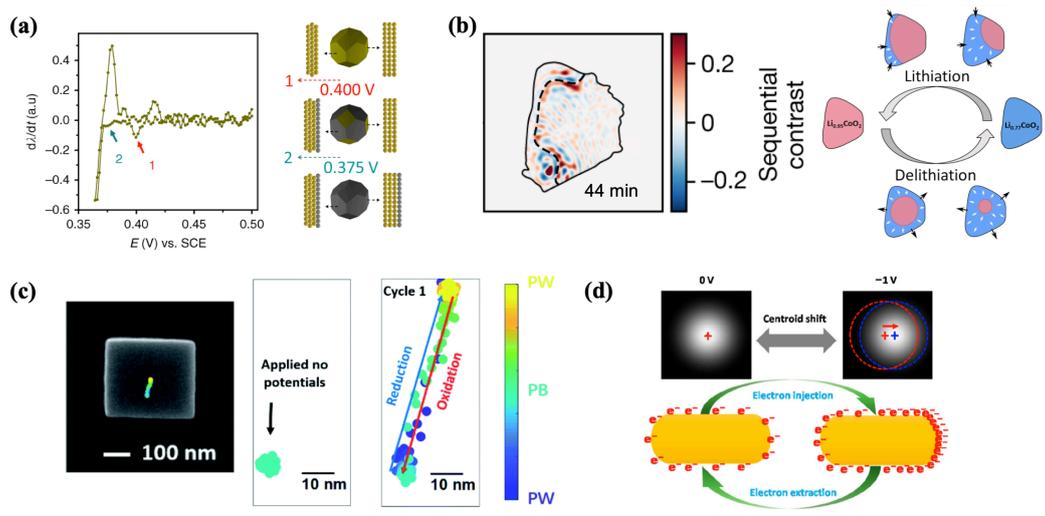

**FIGURE 4**

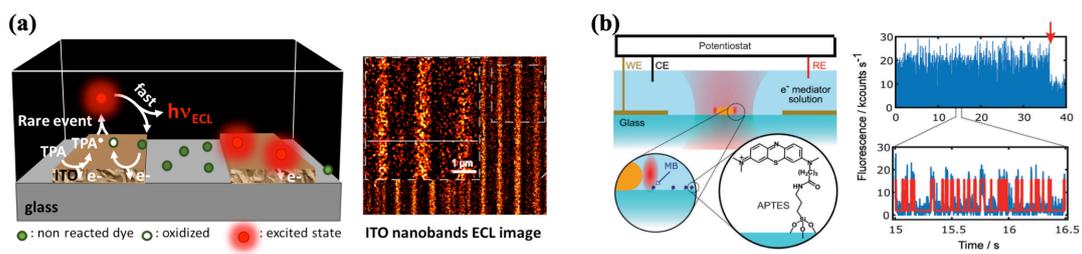

**FIGURE 5**